# Relevance of a magnetic moment distribution and scaling law methods to study the magnetic behavior of antiferromagnetic nanoparticles

N. J. O. Silva[1*], V. S. Amaral[1] and L. D. Carlos[1]

[1]Departamento de Física and CICECO, Universidade de Aveiro, 3810 -193 Aveiro, Portugal

In antiferromagnetic nanoparticles magnetization the linear component, $\chi_{AF}H$ superposed to the saturation one usually complicates the fit of experimental data. We present a method based on scaling laws to determine the variation of $\chi_{AF}$ with temperature and to find the temperature dependence of the average magnetic moment $<\mu>$, without any assumption on both the magnetization dependence on field or the moment distribution function, whose relevance can also be estimated. We have applied this method to ferritin and found that $<\mu>$ decreases with increasing temperature and that a distribution function cannot be ignored. The fit with Langevin magnetization law and lognormal moment distribution functions yielded parameters close to those estimated with the scaling method. We also show that in general if the distribution is ignored, and a single particle moment $\mu_p$ is used, $\mu_p$ presents an artificial systematic increase with temperature. This calls the attention to the necessity of evaluating the effect of a size distribution before concluding about the physical nature of the parameters variation.

Magnetic nanoparticles are of great interest due to their application in high density magnetic storage media, emergent applications in biomedicine as magnetic cell sorting and magnetic fluid hyperthermia. Finite-size and surface effects dominate the magnetic properties as size decreases, leading to unusual properties, distinct from bulk material. Superparamagnetism is a major finite-size effect, where the magnetic anisotropy constitutes an energy barrier for magnetization reversal, that depends on the volume as $E=KV$ (where $K$ is an effective anisotropy constant) and the frequency of thermal activated reversals crossing this barrier is $f=f_0 e^{-E/kT}$ where $f_0$ is an "attempt frequency" [1]. Depending on the particles volume and observation time, the particles are blocked below a characteristic temperature, $T_B$. Above $T_B$ the moments are essentially free to rotate and the $M(H)$ curve can be described with the well known Langevin function, $L(x)$:

$$M(H,T)=m_0[\coth(\mu_p H/kT)-1/(\mu_p H/kT)]$$
$$=m_0 L(\mu_p H/kT) \quad (1)$$

where $m_0$ is the saturation magnetization and $\mu_p$ the magnetic moment. $m_0$ is equal to the product of the number of particles per volume or mass (depending on the $M$ dimension), $N$, for $\mu_p$. This constitutes the classical description where the magnetic moment makes a particular angle with the applied field and the energy levels distribution (function of the angle, field and temperature) follows a Boltzmann law.

This discussion can be generalized to antiferromagnetic nanoparticles, where total magnetization has been described as the sum of a saturation component (associated with uncompensated moments) and a linear term [2, 3]:

$$M(H,T)=m_0 L(\mu_p H/kT)+ \chi_{AF} H \quad (2)$$

Deviations to the Langevin behavior occur due to interparticle interactions [4], anisotropy [5] and inhomogeneities such and volume and moments distribution. In nanoparticles systems, volume and moments distributions are ubiquitous, existing even in biological systems as ferritin, where the iron oxide hydroxide ferrihydrite ($FeOOH \cdot nH_2O$) with ~5nm of diam. is wrapped up in a protein template [6]. This calls the attention for the introduction of such distribution in the analysis of magnetic curves [7-10]. However, some works consider equations (1) and (2) as a valid approach to obtain the mean parameters $m_0$ and $\mu_p$, and their temperature variation [2, 3, 11-17]. The values thus obtained are then compared with structure information [2, 11-16] and their temperature dependence is related with microscopic characteristics as spin arrangements [2], thermoinduced magnetization [18] or interpaticle interactions [14] and is then used to estimate the Néel temperature [2, 16, 17] and derive deviations from the Curie law [19]. In the present work, we show that a distribution must be taken in account to derive correct mean parameters (even for narrow distributions). Moreover, we show that using a non distributed function (as equation (1) and (2)) in a distributed-system with an intrinsic constant $\mu_p$ magnetization curve an increase of $\mu_p$ with temperature is systematically obtained. This dependence is similar to that found in literature for several systems [2, 14-17, 20], and so, major doubt arises on the interpretation of microscopic characteristics and susceptibility behavior based just on that temperature dependence. Furthermore, as the choice of a distribution function is so critical to obtain a proper parameter temperature variation we present a new method to separate the antiferromagnetic and the superparamagnetic components in antiferromagnetic nanoparticles without considering, at the start, a specific law or distribution function for the superparamagnetic part. For the latter component we also discuss a possible way to distinguish between four different system types, considering the existence of scaling laws: non-distributed and distributed systems with constant $\mu_p$ and with $\mu_p$ varying with temperature. This procedure is applied to the magnetization curves of native horse-spleen ferritin. The parameters estimated from this method are then compared with those obtained from fitting with distributed and non-distributed Langevin functions.

The total magnetization of a superparamagnetic system with magnetic moment distribution can be expressed by:

$$M(H,T) = N \int_{\mu_{min}}^{\mu_{max}} \mu L\left(\frac{\mu H}{kT}\right) f(\mu) d\mu \qquad (3)$$

where $N$ is the particle density, $\mu$ the particle moment and $f(\mu)$ its normalized distribution function. In the magnetic measurements $M(H,T)$ we use as relevant quantity to be distributed the magnetic moment with a log-normal distribution expressed by:

$$f(\mu) = \frac{1}{\mu \cdot s\sqrt{2\pi}} e^{-\frac{\left(\log\left(\frac{\mu}{n}\right)\right)^2}{2s^2}} \qquad (4)$$

The mean particle moment $<\mu>$ is equal to $n\sqrt{w}$ and the standard log-normal deviation $\sigma$ is equal to $n\sqrt{w(w-1)}$, with $w = e^{s^2}$. In ideal superparamagnetic systems, $\mu$ is proportional to the volume and the moment distribution arises only due to a volume distribution. In that case it is possible to consider volume distributions instead of moments distribution. However, in real systems, surface disorder, frustration and spin canting may contribute to moment distributions distinct from volume ones [21].

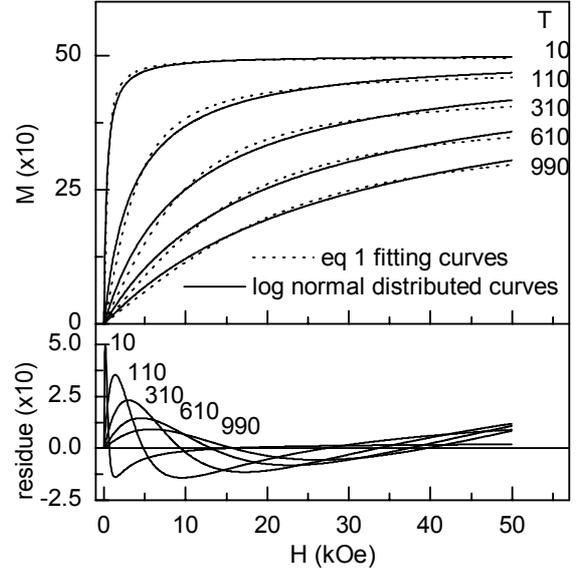

Fig. 1. Curves simulated with eq 3 using a lognormal function for different temperatures, fitted with a non distributed function (eq. 1) (above) and fit residues (below).

To show the relevance of the use of a distribution, we have generated several sets of curves using equation (3), for 10<T<1000, 0<H<50000, with Gaussian and log-normal distributions, different standard deviations, $\sigma$, and $<\mu>$, considering $N=1$. Each set of temperatures was then taken as data from a superparamagnetic system and fitted with the non-distributed Langevin law given by eq. (1). These fits appear quite good, even when using log-normal distribution with a standard log-normal deviation: $\sigma = 1$ (Fig.1). Naturally, as $\sigma$ decreases the differences became smaller. However, a systematic deviation curve shape with two zeros that decreases with temperature is noteworthy (Fig. 1, below). Similar systematic oscillatory fit deviations can be observed in real $M(H,T)$ curves fitted with eq. (1) [11, 13] and eq. (2) [14, 15, 17, 20]. Using the Gaussian distribution, in the limit case of $\sigma = <\mu>/2$ the differences are of the order of 3%. In the low field Curie region, the deviation is higher and implies miss evaluation of $m_0$ and $\mu_p$, as already stated by Harris et al [20]. These authors suggested linear fits to the asymptotic low and high field regions as a way to avoid errors introduced by a non-distributed Langevin fit. However, there is an intrinsic difference between the Curie law ($M/H=C/T$) of a distributed and non-distributed Langevin function: in the later function, the Curie constant is $C = m_0 \mu_p = N \mu_p^2$, while in the former function $C$ depends on the standard deviation: $C=N<\mu^2>=N(<\mu>^2 + \sigma^2)$. Thus, the error on the evaluation of $<\mu>$ using the non-distributed Curie law on a distributed system is of the order of $\sigma/<\mu>$.

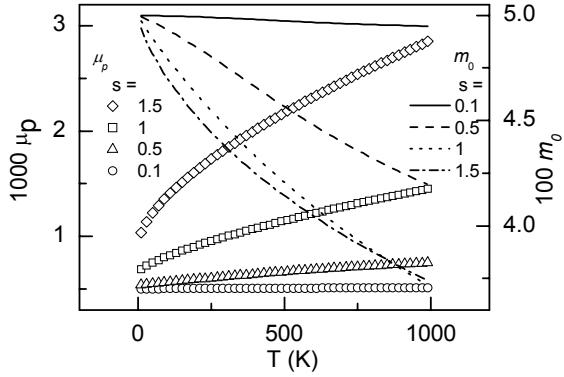

Fig. 2 Fit parameters $\mu_p$ and $m_0$ as a function of temperature for different standard log-normal deviations.

The eq.(1) fit parameters $m_0$ and $\mu_p$ were found to scale with $T/<\mu>H$. In real measurements, the magnetization curves are obtained for different temperatures in a given field range and so, the parameter of interest to study the variation of $m_0$ and $\mu_p$ is T, as shown in Fig 2. The shape of $m_0(T)$ and $\mu_p(T)$ and its relative variation depend on the distribution function, the standard deviation and on the $<\mu>.H_{max}/T$ observation window. General features are the surprisingly low parameters error bar (less than 1%) and the systematic decrease of $m_0$ and increase of $\mu_p$ with temperature. As $\sigma$ increases, the relative variation of $m_0$ and $\mu_p$ becomes more pronounced and for a temperature range of one order of magnitude is greater than 10% for $\sigma >0.4$ (Fig. 2). As temperature reduces $m_0$ and $\mu_p$ tend to the correct values $N<\mu>$ and $<\mu>$, respectively. Such effect of deficient distribution is also observed comparing Gaussian, log-normal and square distributed curves.

The type of $m_0$ variation plotted in Fig 2 was found by several authors fitting eq (2) to ferritin [2], ferrihydrite [15, 16] and NiO [2, 17] magnetization curves and was tentatively associated to a surface moments intrinsic behavior. Our results show that care must be taken to ensure that such an $m_0$ variation has physical meaning and does not come from ignoring a magnetic moment distribution. At the same time, the $\mu_p$ increase is observed in ferritin (see Fig. 6), artificial ferritin with different core mean sizes [20] and ferrihydrite particles [15, 16]. This apparent temperature assisted onset of magnetic moments was associated with weaker exchange, strong radial anisotropies, frustration, multiple sublattices (ref. 20 and references there in) and interparticle interactions [15]. Recent work interprets this anomalous behavior as dynamic thermoinduced magnetization [18]. Despite the possible presence of all these features in the referred systems, our calculations show that the existence of a $\mu$ distribution leads to an analogous $\mu_p$ temperature variation that must be carefully analyzed. A closer look reveals that the reported stronger variations take place in powder ferrihydrite samples [15] and in the smaller artificial ferritins [20], where an higher volume and thus $\mu$ distribution are likely to occur. The simultaneous increase of $\mu_p$ and decrease of $m_0$ is, in fact, contradictory and would imply a strong decrease of the particle density $N$, which has no physical ground.

At this point we have shown how critical the use of an accurate distribution function can be to derive the correct parameters temperature variation. Thus, the question of distinguishing between an intrinsic parameter variation and a deficient fitting function situation becomes critical. Hence, a method to derive both qualitative information about the system and an estimation of the parameters value before a compromise with fit functions is of major interest. In the case of the antiferromagnetic nanoparticles two other problems arise. Firstly, these systems have an antiferromagnetic susceptibility component, which is difficult to separate from the superparamagnetic part especially if the latter is far from saturation. Moreover, the departure from saturation depends on the temperature and an high field linear fit gives $\chi_{AF}$ for excess and successively far from the accurate value as the temperature rises. The variation of $\chi_{AF}$ with temperature has not been modelled yet; although, a $T^{-0.5}$ variation has been proposed by Gilles et al. [22]. Secondly, the superparamagnetic component can be modeled with the Langevin function, as expressed in equations 2 and 3, a Langevin function with a modified $m_0$ factor [20] or with the "random magnetic orientation" function derived by Néel (Ref [22, 23] and references there in). Both Langevin and Néel functions have the same asymptotic behavior but differ in the intermediate field zone, with the latter function saturating at lower fields. In the following paragraphs we present a method to derive qualitative information about antiferromagnetic particles parameters variation, we discuss the limits of application, the types of systems which can be discerned and we apply this method to the case of native horse-spleen ferritin, obtained from Sigma Chemical Company and prepared in samples accordingly to ref [2]. Sample magnetic field dependence up to 50 kOe at several temperatures (30-250K) was measured with a Quantum Design SQUID magnetometer after field cooling (5kOe).

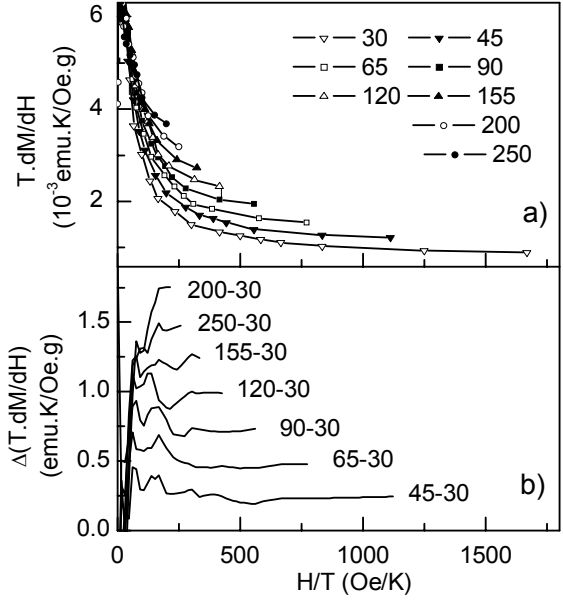

Fig. 3 a) Representation of $(\partial M/\partial H)T$ as a function of $H/T$ for horse spleen ferritin; b) difference between each of the above curve and the T=30K curve as a function of $H/T$.

In the case of antiferromagnetic nanoparticles with $<\mu>$ constant with temperature, the supeparamagnetic component scales with $H/T$, independently of the distribution function or the above mentioned laws that describe the system. The $\chi_{AF} H$ component would scale with $H/T$ only if $\chi_{AF}$ obeyed to a Curie law. The first derivative of the magnetization with respect to the field multiplied by temperature has a component that collapses in an $H/T$ scale and another component associated with from $\chi_{AF}$ in accordance with:

$$\frac{\partial M}{\partial H}T = F\left(\frac{H}{T}\right) + \chi_{AF}T \qquad (5)$$

where $F$ is an unknown function of $H/T$. Thus, if $\chi_{AF}$ does not follow a Curie law, a constant difference between the magnetization data obtained at different temperatures will appear in an $(\partial M/\partial H)T$ vs. $H/T$ plot, as shown in Fig. 3a) for ferritin. The increment of $\chi_{AF}T$ in relation to a given temperature $T_0$, can be evaluated in a $(\partial M/\partial H)T - (\partial M/\partial H)T_0$ representation. At this point, the accuracy of supposing $<\mu>$ constant with temperature can be checked as the constancy of $(\partial M/\partial H)T - (\partial M/\partial H)T_0$ with temperature. In the case of native horse-spleen ferritin we chose the lower studied temperature 30 K as $T_0$ and we observe a region where the curves can be considered constant with $H/T$ and another region where significant variations occur (Fig. 3b)). We should note that even if $F$ is not just a function of $H/T$, $(\partial M/\partial H)T - (\partial M/\partial H)T_0$ tends to a constant value as $F$ tends to saturation and the plot $(\partial M/\partial H)T - (\partial M/\partial H)T_0$ vs. $H/T$ can be used to observe how far the curves are from that condition. To have further information, $\chi_{AF}$ at the lowest temperature must be determined. In ferritin, $\chi_{AF}(T)$ can be determined considering the regions where $(\partial M/\partial H)T - (\partial M/\partial H)T_0$ can be considered constant (Fig. 3b)) and a value for $\chi_{AF}$ (T=30). That value can be obtained from the extrapolation to zero of $\partial M/\partial H$ as a function of $T/H$ and was estimated as $\chi_{AF}$ (T=30)=2.6 × 10$^{-5}$ emu/Oe.g. Then, the superparamagnetic component of ferritin can be obtained using and subtracting $\chi_{AF}.H$ to the total magnetization (Fig. 4a)). As noticed in the variation of $(\partial M/\partial H)T - (\partial M/\partial H)T_0$, the curves do not superimpose in $H/T$ and thus one concludes that $<\mu>$ should vary with temperature. Moreover, as the curves saturate successively at higher $H/T$ as the temperature of measurement is higher we can qualitatively infer that $<\mu>$ decreases with temperature.

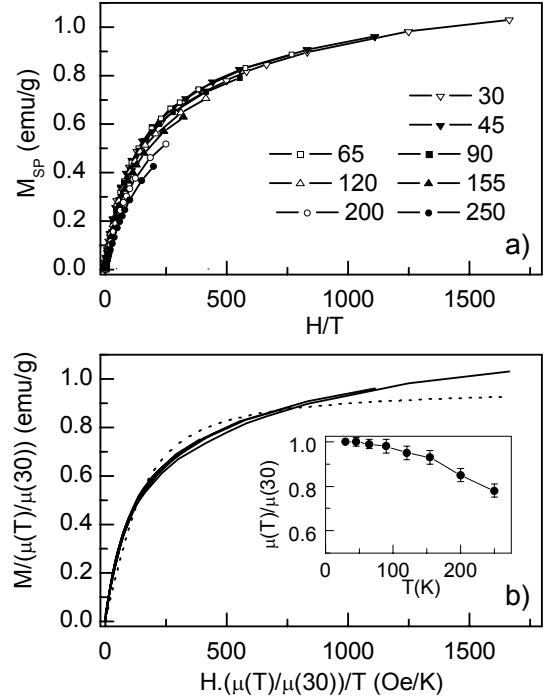

Fig. 4 a) superparamagnetic (saturation) component as a function of $H/T$; b) saturation component in a the scaling axes $M/(\mu(T)/\mu(30))$ vs. $H(\mu(T)/\mu(30))/T$. The non-distributed Langevin fit is shown as dotted line. Inset shows the relative μ temperature variation.

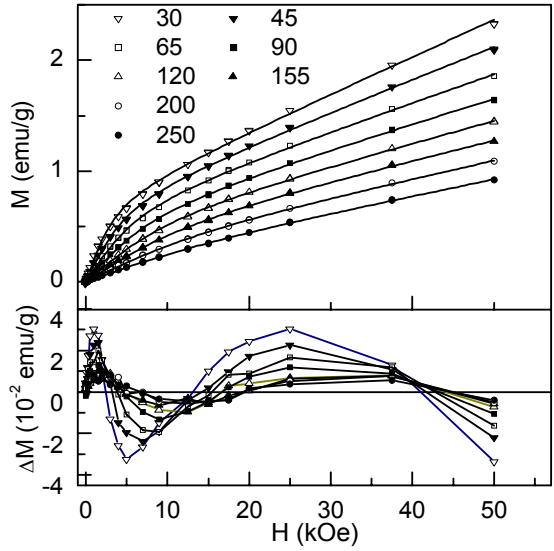

Fig 5 Magnetization of ferritin as a function of applied field at the indicated temperatures. Solid lines represent the fits to the non-distributed Langevin law (eq. 2). Below: fit residues.

So, without any assumption of a particular function or distribution, we show that there is a set of $\chi_{AF}$ values that give left a superparamagnetic component where $<\mu>$ decreases with temperature. In a general case, if the superparamagnetic component curves scale, a single fit to all temperatures can be performed and several laws and distribution functions can be tested, avoiding $\chi_{AF}$ and knowing beforehand the qualitatively variation of $<\mu>$, $N$ and other parameters with temperature. Moreover, the value of $\chi_{AF}(T_0)$ can be refined imposing the variation derived from the $(\partial M/\partial H)T - (\partial M/\partial H)T_0$ and searching for the $\chi_{AF}(T_0)$ that minimizes the differences between the superparamagnetic component in the $H/T$ scale. In the cases of the superparamagnetic component of an antiferromagnetic or in ferromagnetic systems where $<\mu>$ is temperature dependent we can still distinguish between situations where the distribution of $\mu$ can or cannot be ignored, before doing any fit. Whenever is possible to ignore the distribution of $\mu$, there is a scaling factor for each curve such that dividing the magnetization and multiplying the $H/T$ scales all curves. In other words, if the $\mu$ distribution can be ignored, it is possible to find a $M/\mu(T)$ vs. $H.\mu(T)/T$ scaling plot. This happens in two limit situations: whenever the distribution is narrow or the temperature variation of $\mu$ with temperature is small compared with the distribution deviation. In order to find that scale plot, the lower temperature curve can be set as a reference and the ratio $\mu(T)/\mu(T_0)$ derived. In our ferritin results there are no such scale factors and thus a distribution function cannot be ignored. However, ferritin approaches the case where variation of $<\mu>$ with temperature is small compared with the distribution deviation. At the same time, as the $\mu$ distribution influences just the low and middle $H/T$ region, while the high $H/T$ limit is characterized by $N.<\mu>$, the $<\mu(T)>/<\mu(T_0)>$ ratio can be estimated by searching for scaling in that limit. Doing this for ferritin, we obtain a decrease ratio of 0.78±0.03 when the temperature increases from 30 to 250K. In summary, without knowing the particular distribution function, this method gives information about the $\chi_{AF}$ temperature dependence and the existence of a $\mu$ constant or temperature dependent. The absolute scale of $\chi_{AF}$ and $<\mu(T)>$ are determined by $\chi_{AF}$ and $\mu$ at the reference temperature $T_0$.

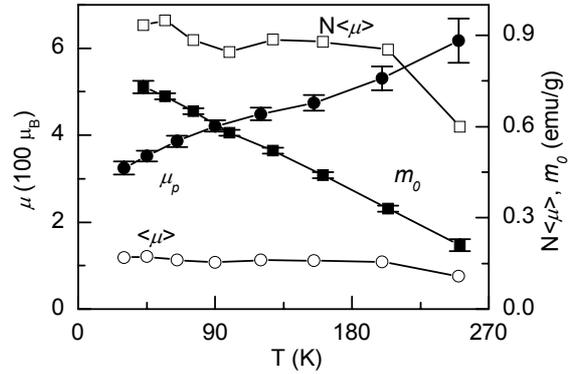

Fig. 6 Parameters $\mu_p$ and $m_0$ obtained with a non-distributed Langevin function (eq. 1) compared with the correspondent parameters $<\mu>$ and $N<\mu>$ obtained with a log-normal distributed Langevin function (eq. 3) (fitting all curves).

Further data analysis is therefore enlightened by this apriori information about the characteristics of the magnetization curves of ferritin, namely that the a distribution function cannot be ignored and that $\mu(T)$ decrease with temperature. Fitting the ferritin magnetization curves with a non-distributed Langevin function, eq.(2) yielded systematic residue (Fig. 5) and $m_0$ and $\mu_p$ decreasing and increasing with temperature, respectively (Fig. 6), in accordance with earlier results [2] and resembling the previous discussed variation found fitting eq.(2) to distributed data (Fig. 2). Fitting each magnetization curve using a log-normal function in eq (3) we find that the residues are of the order of data scattering. However, for T>155, where the curvature is small, the fit does not converge. Using the "random magnetic orientation" function (Ref [22, 23] and references there in) together with the lognormal distribution we obtain non-linear least square values of about 5% higher than using Langevin and lognormal distribution functions. As the particles density $N$ obtained in these individual fits remain essentially constant with temperature, we performed a global fit imposing the same $N$ for all temperatures. The $<\mu>$ and $N<\mu>$ thus obtained are compared with $m_0$ and $\mu_p$ in Fig. 6. In Fig 7 we observe that $\chi_{AF}$ decreases with temperature according with the variation expected form the scale method analysis (where the absolute scale was set considering the value obtained from this fit at T=30 K). In fact, the $\chi_{AF}$ variation obtained from the scale method is even smoother that the obtained from the fit, especially at higher temperatures, where individual fits did not converge. The same is observed for $<\mu(T)>$: the variation is of the order of the value obtained from the scale method (assuming $\mu(30)$ obtained from the fit) and scale method gives a smoother variation at higher temperatures. This type of variation was already found by Gilles et al., using a log-normal distribution function and a 'random magnetic orientation' function [22]. The value of $\mu(30)$ is about 5 times lower than the value obtained using a non-distributed Langevin function (eq.(2)) and about 0.8 of the value obtained in ref [22], corresponding to a mean number of fully uncompensated $Fe^{3+}$ ions, $N_{un}$ of 23. The number of $Fe^{3+}$ ions involved in the superparamagnetism is obviously grater and a range between fully compensated and fully uncompensated configurations is expected. As the mean horse-spleen ferritin core has a total number of Fe ions, $N_t$, of about 2000-3000 [6], $N_{un}$ is of the order of $N_t{}^p$ with $1/2<p<1/3$, which suggests that the uncompensated spin are not only at the surface but also randomly distributed through the volume. The lognormal deviation $\sigma$ varies from a 0.9 at 30K to 1.3 at 65K and to 1.0 at 250K. These values are not compatible with the assumption that the number of uncompensated moments is a fixed power $p$ of the volume and that the uncompensated moments distribution is just a consequence of volume distribution. In fact, $\sigma = 1$ implies the existence of particles with a maximum number of fully uncompensated $Fe^{3+}$ ions of about 100 ions, which means 10 000 total ions with $p=1/2$, while the ferritin maximum Fe loading is about 5000 ions. In other words, there are particles with the same volume that have different magnetic moments, probably due to different degrees of structure/magnetic disorder.

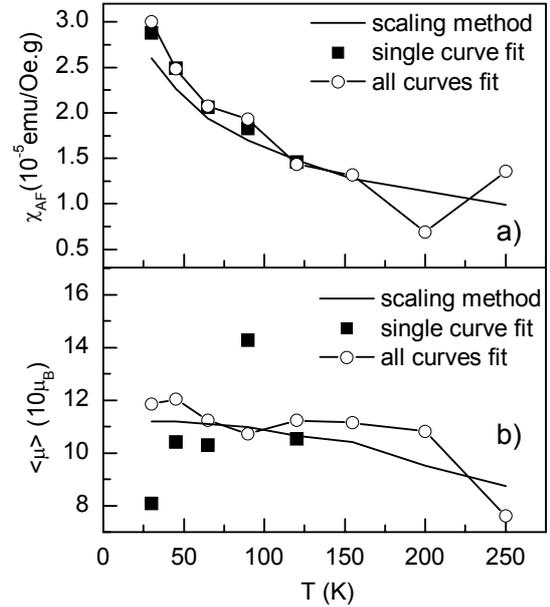

Fig 7 eq. 3 fit parameters $\chi_{AF}$ (above) and $<\mu>$ (below) obtained fitting each curve separately, fitting all curves imposing the same particle density in all temperatures, compared with the values predicted by the scaling method.

The magnetization of antiferromagnetic nanoparticles has an extra linear component, $\chi_{AF}$, merged with a saturation one, which brings additional problems to perform the fit. As the choice of using a distribution function is critical and, at the same time, the existence of exotic behavior, as the $\mu_p$ increase, can not be withdrawn, it is extremely important to get information about the characteristics of the system before doing any fit. In this paper, we present a new method based on scaling laws to determine the variation $\chi_{AF}$ with temperature. Together with the estimation of $\chi_{AF}$ at the lowest temperature it is also possible to determine either $<\mu>$ is or is not constant with temperature and an estimation of its variation. If $<\mu>$ depends on the temperature, we can also determine if the system can be described without a distribution function or if the variation of $<\mu>$ with temperature is small comparing with the distribution deviation. We have applied this method for ferritin and we found that $<\mu>$ decreases with temperature and that a distribution function can not be ignored. The subsequent fit with Langevin and lognormal functions yielded parameters of the order of those estimated with the scaling method. However, the variation of the latter parameters is even smoother than that obtained from the fit, which can be regarded as an indication of smaller errors. The moment distribution is not compatible with the narrow size distribution characteristic of ferritin. Thus, we are lead to conclude about the existence of important intra-particle magnetic disorder as the source of moment distribution or, in alternative, that the Langevin and "random magnetic orientation" functions are not suitably describing ferritin. We have also studied some consequences of using a non distributed Langevin function to fit the magnetization curve of an intrinsic log-normal distributed system. First, a systematic '~ 'shaped difference, similar to several found in literature is observed. Secondly, the fit parameters thus obtained $m_0$ and $\mu_p$, have low error bar and diverge from the real values as the $H/T$ observation window decreases. At last, $m_0$ decreases while $\mu_p$ increases with temperature, like previously found in real systems as ferritin [2], ferrihydrite [15, 16], and NiO [**Error! Reference source not found.**] and attributed to fundamental behavior as surface effects or thermoinduced magnetization [18]. This calls the attention to the necessity of evaluate the effect of distributions before conclude about the physical nature of the parameters variation.

Acknowledgments

The authors acknowledge M. S. Reis for helpful discussions. This work was supported by FEDER and FCT research project POCTI/P/CTM/33653/00 and POCTI/CTM/46780/02. N. J. O. Silva acknowledges a scholarship by FCT (SFRH/BD/10383/2002). The possibility of performing the SQUID magnetic measurements in the Materials Institute of the University of Porto (IFIMUP) is gratefully acknowledged.